\documentclass[conference]{IEEEtran}
\IEEEoverridecommandlockouts
\usepackage{cite}
\usepackage{amsmath,amssymb,amsfonts}
\usepackage{algorithmic}
\usepackage{graphicx}
\usepackage{textcomp}
\usepackage{xcolor}
\usepackage{setspace}
\usepackage{subfigure}
\usepackage{amstext}
\usepackage{array}
\usepackage{algorithm}
\def\BibTeX{{\rm B\kern-.05em{\sc i\kern-.025em b}\kern-.08em
    T\kern-.1667em\lower.7ex\hbox{E}\kern-.125emX}}
\begin{document}

\title{Optimal Real-time Communication in 6G Ultra-Massive V2X Mobile Networks\\
}
\author{\IEEEauthorblockN{ He Huang$^{1,3}$, Zilong Liu$^{2}$, Zeping Sui$^{2}$, Wei Huang$^{3}$, Md. Noor-A-Rahim$^{4}$, Haishi Wang$^{1}$, Zhiheng Hu$^{1}$}
$^{1}$College of Communication Engineering, Chengdu University of Information Technology, China\\
$^{2}$School of Computer Science and Electronics Engineering, University of Essex, UK\\
$^{3}$Intelligent Interconnected Systems Laboratory of Anhui Province, Hefei University of Technology, China\\
$^{4}$School of Computer Science and Information Technology, University College Cork, Ireland\\
Email: huanghe@cuit.edu.cn; zilong.liu@essex.ac.uk; z.sui@essex.ac.uk; huangwei@hfut.edu.cn; \\m.rahim@cs.ucc.ie; whs@cuit.edu.cn; hzh@cuit.edu.cn
\thanks{Corresponding author: whs@cuit.edu.cn.}
\thanks{This work was supported in part by Fundamental Research Funds for the Central Universities of China (Grant No.PA2024GDSK0114 and No.JZ2024HGTG0311),  National Natural Science Foundation of China (Grant No.62371180), Open Research Project of the State Key Laboratory of Industrial Control Technology, China (Grant No. ICT2025B23), Sichuan Science and Technology Program (Grant No.2025HJRC0023), Research Fund of Talent Introduction of Chengdu University of Information and Technology (Grant No.376613).}
}
\maketitle

\begin{abstract}
This paper introduces a novel cooperative vehicular communication algorithm tailored for future 6G ultra-massive vehicle-to-everything (V2X) networks leveraging integrated space-air-ground communication systems. Specifically, we address the challenge of real-time information exchange among rapidly moving vehicles. We demonstrate the existence of an upper bound on channel capacity given a fixed number of relays, and propose a low-complexity relay selection heuristic algorithm. Simulation results verify that our proposed algorithm achieves superior channel capacities compared to existing cooperative vehicular communication approaches. 
\end{abstract}


\section{Introduction}
\subsection{Background}
The rapid evolution of wireless communications has accelerated the global commercialization and deployment of fifth-generation (5G) networks. Concurrently, research on the sixth generation (6G) of wireless communications has intensified, promising revolutionary enhancements in communication quality, service versatility, and the full realization of integrated space-air-ground communications [1, 2]. Compared to 5G, 6G is expected to yield significantly enhanced channel capacity, reduced outage probabilities, improved security, and enlarged service coverage. Moreover, 6G is deemed to integrate the intelligence, flexibility, and ubiquitous computing for massive Internet of Things (IoT) ecosystems [3].

Among various application scenarios envisioned for 6G, the Cooperative Internet of Vehicles (CIoV) has attracted significant research attention in recent years. Technically, CIoV involves extensive communications among highly dynamic vehicles and requires advanced technologies such as mobile edge computing (MEC), artificial intelligence (AI), and comprehensive space-air-ground integration [4]. Extensive research efforts have been dedicated to addressing challenges in large-scale vehicular networks, including cooperative relay selection, AI-driven vehicular task coordination, and intelligent edge computing algorithms. For instance, a distributed cooperative caching strategy was proposed in [5] to maximize device-to-device (D2D) traffic in edge networks. Similarly, [6] developed an energy allocation scheme to optimize the long-term energy efficiency under certain communication and computation constraints. Furthermore, distributed state estimation frameworks utilizing 5G technology have been explored to enhance smart grid performance [7]. In short, optimizing real-time communications in future ultra-massive V2X networks will represent a crucial research frontier.

\subsection{Motivations and Contributions}
Against the above background, the following two research problems remain largely open in ultra-massive V2X communication scenarios:

\begin{enumerate}
\item Limitations of Fixed Base Stations: Traditional fixed base station deployments are inadequate for supporting reliable vehicular communications due to rapid vehicular mobility, resulting in coverage gaps and unstable transmission quality. Hence, adaptive relays selection is essential to ensure robust connectivity among highly mobile vehicles.
\item Real-Time Information Dissemination: The current vehicular traffic information primarily relies on satellite-based sources (e.g., Google Maps or Baidu Maps), which do not reflect instantaneous road conditions. Despite existing standards such as IEEE 1609.X, IEEE 802.11p, and ETSI ITS-G5, real-time message dissemination remains underdeveloped. Consequently, novel real-time data processing frameworks must be designed to enable immediate vehicular response to dynamic traffic scenarios.
\end{enumerate}

Driven by the above challenges, the main contributions of this work are summarized as follows:

\begin{enumerate}
\item We propose a dynamic and ultra-scalable V2X framework, by utilizing cooperative relays consisting of vehicles, unmanned aerial vehicles (UAVs), satellites, and other mobile IoT nodes. We will show that the proposed framework facilitates the fulfillment of  stringent 6G requirements such as ultra-fast communications, precision positioning, autonomous driving, and real-time high-definition mapping.

\item We prove that the channel capacity achieves an upper bound for a fixed number of relays. Leveraging inequality theory, we derive an efficient, low-complexity multi-relay selection algorithm and we reveal that if we allocate higher power for device vehicles, better channel capacity will be achieved. Besides, we demonstrate that larger number of relays do not ensure channel capacity improved, so we shall compute suitable number of relays. Simulation results demonstrate a notable enhancement in channel capacity—achieving improvements of approximately 0.8-2.5 Kbps over existing V2X solutions.
\end{enumerate}

\textbf{\emph{Notations}:}
Let ${\Re ^{m \times n}}$ denote the space of $m \times n$-dimensional real matrices, and let ${{\bf{K}}^{\bf{T}}}$ represent the transpose of matrix ${\bf{K}}$. The operator $\circ$ denotes the Hadamard (element-wise) product, and $\left| . \right|$ represents the absolute value. The notation ${R_i} \to {R_j}$ indicates that relay ${R_i}$ transmits a signal to relay ${R_j}$. Finally, $\left\{ {{R_L}} \right\}$ denotes a subset of relays containing $L$ relays.

\begin{figure}[htbp]
  \centering
  \includegraphics[width=3.5 in]{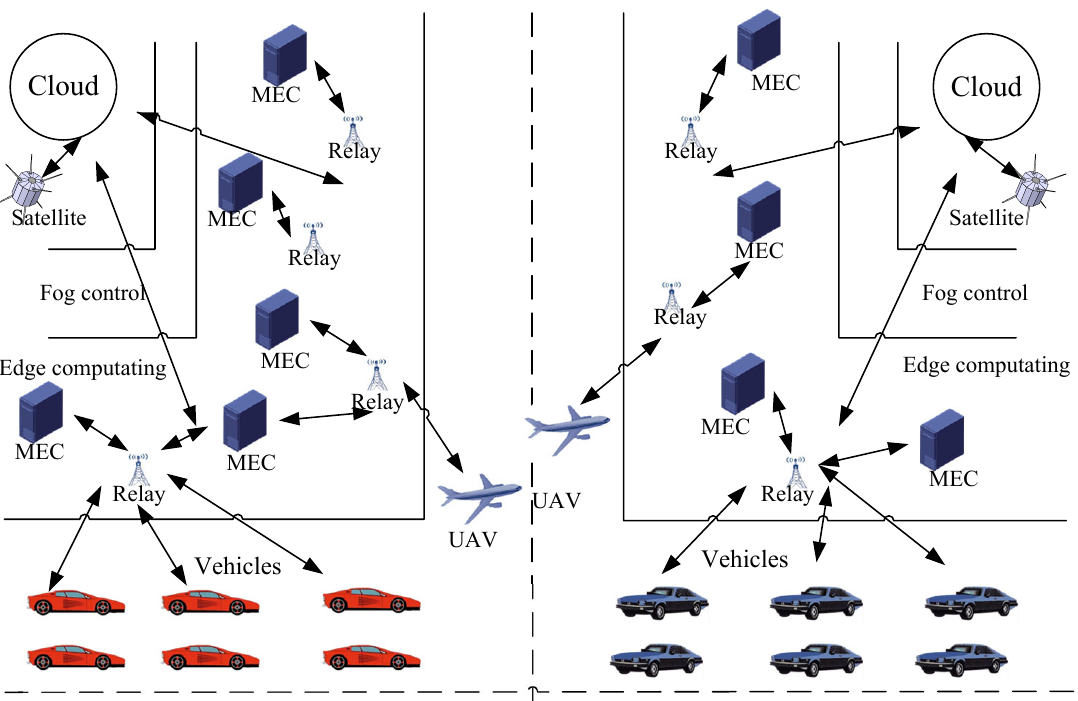}
  \caption{6G ultra massive V2X ubiquitous networks}
  \label{Fig.1}
\end{figure}

\section{Proposed V2X System Model}

In Figure 1, similar to the communication scenarios in ref. [2, 3, 4], we consider a 6G ultra-massive air-space-ground V2X ubiquitous network comprising a large number of vehicles, UAVs, satellites, fixed base stations, and MEC equipments. To investigate ultra-massive V2X mobile networking, we will distinguish these nodes with different channels coefficients in Section five. In addition, we focus on the study of optimal communication in cooperative D2D networks. Following [5] and [10], we assume that each communication device operates in half-duplex mode with a single antenna. We also neglect the impact of multiple antennas and consider that D2D links have distinct channel coefficients for different device nodes.

\begin{figure}[htbp]
  \centering
  \includegraphics[width=3.5 in]{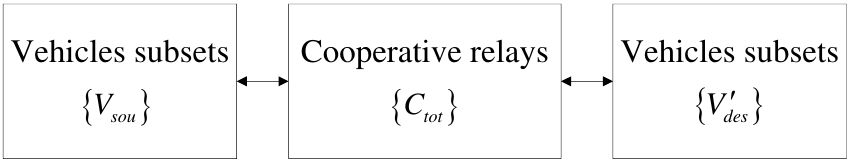}
  \caption{Communications of multiple ultra large scale subsets}
  \label{Fig.1}
\end{figure}

We assume that there are ${k_1}({k_1} \in {{\bf{{\rm N}}}^ * })$ vehicles in the source set $\left\{ {{V_{sou}}} \right\}$, defined by $\left\{ {{V_{sou}}} \right\} = ({v_1},{v_2},...,{v_{{k_1}}})$. We also define the total relay set  $\left\{ {{C_{tot}}} \right\}$, consisting of ${N_{tot}}$ relays. This set includes several subsets: $\left\{ {{V_c}} \right\}$ (vehicle relays), $\left\{ {{\rm{UAV}}} \right\}$ (UAV relays), $\left\{ {{\rm{MOB}}} \right\}$ (mobile relays, such as handsets, laptops, and small IoT devices), $\left\{ {{\rm{REL}}} \right\}$ (fixed base station relays). We assume that the vehicles in the source set $\left\{ {{V_{sou}}} \right\}$ and the destination set  $\left\{ {{{V'}_{des}}} \right\}$ are completely distinct from the vehicles in the cooperative relay set $\left\{ {{C_{tot}}} \right\}$.

\section{Optimal communication in ultra large scale V2X networks}

In Fig. 2, for a source set $\left\{ {{V_{sou}}} \right\}$ with $k_1$ vehicles and a destination set $\left\{ {{{V'}_{des}}} \right\}$ with $k_2$ vehicles, let  $\forall {v_{{k_i}}}$ be a vehicle in $\left\{ {{V_{sou}}} \right\}$ where $1 \le k_i \le k_1$, and $\forall {{v'}_{{k_j}}}$ be a vehicle in $\left\{ {{{V'}_{des}}} \right\}$ where $1 \le k_j \le k_2$. When ${{{v'}_{{k_j}}}}$ sends a signal to ${v_{{k_i}}}$ (${{v'}_{{k_j}}} \to {v_{{k_i}}}$), an arbitrary set of $L$ relays is selected from the total set of $N_{tot}$ relays, denoted by $\left\{ {{R_L}} \right\} = ({R_1},{R_2}, \ldots ,{R_L})$, where $\left\{ {{R_L}} \right\} \subseteq \left\{ {{C_s}} \right\}$ and $L \in \mathbb{N}^*$. We assume the transmission power of relay $R_m$ is $Q_{R_m}$ for ${R_m} \in \left\{ {{R_L}} \right\}$, and the transmission powers of the relay subset $\left\{ {{R_L}} \right\}$ are given by $\left\{ {{Q_{{R_L}}}} \right\} = ({Q_{{R_1}}},{Q_{{R_2}}}, \ldots ,{Q_{{R_L}}})$. In time slot $t$, the sensing information of vehicle ${{v'}_{{k_j}}}$ is ${y_{{{v'}_{{k_j}}}}}$, its transmission power is ${Q_{{{v'}_{{k_j}}}}}$, the channel coefficient between ${{{v'}_{{k_j}}}}$ and relay $R_m$ is ${h_{{{v'}_{{k_j}}}{R_m}}}$, and the noise at relay $R_m$ is modeled as additive white Gaussian noise (AWGN) with variance $\sigma _{{R_m}}^2$. We assume dimension of $n$ for the following vectors:
${\bf{y}}_{{{{\bf{v'}}}_{{{\bf{k}}_{\bf{j}}}}} \to {{\bf{R}}_{\bf{m}}}}^{} \in {\Re ^{1 \times n}}$,
${{{\bf{v'}}}_{{{\bf{k}}_{\bf{j}}}}} \in {\Re ^{1 \times n}}$,
${{\bf{R}}_{\bf{m}}} \in {\Re ^{1 \times n}}$,
${{\bf{Q}}_{{{{\bf{v'}}}_{{{\bf{k}}_{\bf{j}}}}}}} \in {\Re ^{1 \times n}}$,
${{\bf{h}}_{{{{\bf{v'}}}_{{{\bf{k}}_{\bf{j}}}}}{{\bf{R}}_{\bf{m}}}}} \in {\Re ^{1 \times n}}$,
${{\bf{y}}_{{{{\bf{v'}}}_{{{\bf{k}}_{\bf{j}}}}}}} \in {\Re ^{1 \times n}}$, and
${\bf{\sigma }}_{{{\bf{R}}_{\bf{m}}}}^{\bf{2}} \in {\Re ^{1 \times n}}$.

In the first time slot, the signal received at relay $R_m$ from vehicle ${{v'}_{{k_j}}}$ is expressed as:
\begin{equation}
\begin{aligned}
{\bf{y}}_{{{{\bf{v'}}}_{{{\bf{k}}_{\bf{j}}}}} \to {{\bf{R}}_{\bf{m}}}}^{} = \sqrt {{{\bf{Q}}_{{{{\bf{v'}}}_{{{\bf{k}}_{\bf{j}}}}}}}}  \circ {{\bf{h}}_{{{{\bf{v'}}}_{{{\bf{k}}_{\bf{j}}}}}{{\bf{R}}_{\bf{m}}}}} \circ {{\bf{y}}_{{{{\bf{v'}}}_{{{\bf{k}}_{\bf{j}}}}}}} + {\bf{\sigma }}_{{{\bf{R}}_{\bf{m}}}}^{\bf{2}}.
\end{aligned}
\end{equation}

If the channel coefficient between relay $R_m$ and vehicle $v_{k_i}$ is $h_{R_mv_{k_i}}$ and the noise at $v_{k_i}$ is AWGN with variance $\sigma_{v_{k_i}}^2$, the received signal in the second time slot is given by:
\begin{equation}
\begin{aligned}
{\bf{y}}_{{{\bf{R}}_{\bf{m}}} \to {{\bf{v}}_{{{\bf{k}}_{\bf{i}}}}}}^{} = \sqrt {{{\bf{Q}}_{{{\bf{R}}_{\bf{m}}}}}}  \circ {{\bf{h}}_{{{\bf{R}}_{\bf{m}}}{{\bf{v}}_{{{\bf{k}}_{\bf{i}}}}}}} \circ {\bf{y}}_{{{{\bf{v'}}}_{{{\bf{k}}_{\bf{j}}}}} \to {{\bf{R}}_{\bf{m}}}}^{} + {\bf{\sigma }}_{{{\bf{v}}_{{{\bf{k}}_{\bf{i}}}}}}^{\bf{2}}.
\end{aligned}
\end{equation}

Substituting (1) into (2), the signal received by $v_{k_i}$ from ${{v'}_{{k_j}}}$ through relay $R_m$ is shown in (3) as ${\bf{y}}_{{{{\bf{v'}}}_{{{\bf{k}}_{\bf{j}}}}} \to {{\bf{v}}_{{{\bf{k}}_{\bf{i}}}}}}^{{{\bf{R}}_{\bf{m}}}}$. When the selected relay subset is $\left\{ {{R_L}} \right\}$, the combined received signal is denoted as ${\bf{y}}_{{{{\bf{v'}}}_{{{\bf{k}}_{\bf{j}}}}} \to {{\bf{v}}_{{{\bf{k}}_{\bf{i}}}}}}^{\left\{ {{{\bf{R}}_{\bf{L}}}} \right\}}$. The corresponding signal-to-noise ratio (SNR) for the transmission from ${{{v'}_{{k_j}}}}$ to ${{v_{{k_i}}}}$ via the relays set $\left\{ {{R_L}} \right\}$ is given by (4).
\begin{figure*}
\begin{equation}
\begin{aligned}
{\bf{y}}_{{{{\bf{v'}}}_{{{\bf{k}}_{\bf{j}}}}} \to {{\bf{v}}_{{{\bf{k}}_{\bf{i}}}}}}^{{{\bf{R}}_{\bf{m}}}}{\rm{ = }}\sqrt {{{\bf{Q}}_{{{\bf{R}}_{\bf{m}}}}}}  \circ {{\bf{h}}_{{{\bf{R}}_{\bf{m}}}{{\bf{v}}_{{{\bf{k}}_{\bf{i}}}}}}} \circ (\sqrt {{{\bf{Q}}_{{{{\bf{v'}}}_{{{\bf{k}}_{\bf{j}}}}}}}}  \circ {{\bf{h}}_{{{{\bf{v'}}}_{{{\bf{k}}_{\bf{j}}}}}{{\bf{R}}_{\bf{m}}}}} \circ {{\bf{y}}_{{{{\bf{v'}}}_{{{\bf{k}}_{\bf{j}}}}}}} + {\bf{\sigma }}_{{{\bf{R}}_{\bf{m}}}}^{\bf{2}}) + {\bf{\sigma }}_{{{\bf{v}}_{{{\bf{k}}_{\bf{i}}}}}}^{\bf{2}}.
\end{aligned}
\end{equation}
\end{figure*}
\begin{figure*}
\begin{equation}
\begin{aligned}
SNR_{{{{\bf{v'}}}_{{{\bf{k}}_{\bf{j}}}}} \to {{\bf{v}}_{{{\bf{k}}_{\bf{i}}}}}}^{\left\{ {{{\bf{R}}_{\bf{L}}}} \right\}} = \frac{{{{\left| {\sqrt {{{\bf{Q}}_{\left\{ {{{\bf{R}}_{\bf{L}}}} \right\}}}}  \circ {{\bf{h}}_{\left\{ {{{\bf{R}}_{\bf{L}}}} \right\}{{\bf{v}}_{{{\bf{k}}_{\bf{i}}}}}}} \circ \sqrt {{{\bf{Q}}_{{{{\bf{v'}}}_{{{\bf{k}}_{\bf{j}}}}}}}}  \circ {{\bf{h}}_{{{{\bf{v'}}}_{{{\bf{k}}_{\bf{j}}}}}\left\{ {{{\bf{R}}_{\bf{L}}}} \right\}}}} \right|}^{\bf{2}}}}}{{{{\left| {\sqrt {{{\bf{Q}}_{\left\{ {{{\bf{R}}_{\bf{L}}}} \right\}}}}  \circ {{\bf{h}}_{\left\{ {{{\bf{R}}_{\bf{L}}}} \right\}{{\bf{v}}_{{{\bf{k}}_{\bf{i}}}}}}} \circ {\bf{\sigma }}_{\left\{ {{{\bf{R}}_{\bf{L}}}} \right\}}^{\bf{2}}} \right|}^{\bf{2}}}{\bf{ + }}{{\left| {{\bf{\sigma }}_{{{\bf{v}}_{{{\bf{k}}_{\bf{i}}}}}}^{\bf{2}}} \right|}^{\bf{2}}}}} \circ {\left| {{{\bf{y}}_{{{{\bf{v'}}}_{{{\bf{k}}_{\bf{j}}}}}}}} \right|^{\bf{2}}}.
\end{aligned}
\end{equation}
\end{figure*}

\textbf{Upper bound of SNR for single vehicle to single vehicle}: Consider $L$ distinct communication paths from vehicle $v'_{k_j}$ to vehicle $v_{k_i}$ via relays $R_1, R_2, \dots, R_L$, respectively. If the noise variance at $v_{k_i}$ is zero, i.e., $\sigma_{v_{k_i}}^2 = 0$, then SNR for the combined path $SNR_{v'_{k_j} \to v_{k_i}}^{\{R_L\}}$ is upper-bounded by the sum of the SNRs for the individual paths $\sum_{w=1}^L SNR_{v'_{k_j} \to v_{k_i}}^{R_w}$.

From (1) and (2), the received signal at $v_{k_i}$ via the relays $\{R_L\}$ is given by (5).
\begin{figure*}
\begin{equation}
y_{v'_{k_j} \to v_{k_i}}^{\{R_L\}} = \sum_{g=1}^L \left( \sqrt{Q_{R_g}} h_{R_g v_{k_i}} \sqrt{Q_{v'_{k_j}}} h_{v'_{k_j} R_g} y_{v'_{k_j}} + \sqrt{Q_{R_g}} h_{R_g v_{k_i}} \sigma_{R_g} \right) + L \sigma_{v_{k_i}}^2.
\end{equation}
\end{figure*}

The SNR for the combined path is expressed as
\begin{equation}
SNR_{{{v'}_{{k_j}}} \to {v_{{k_i}}}}^{\{ {R_L}\} } = \frac{{{{\left( {\sum\limits_{g = 1}^L {\sqrt {{Q_{{R_g}}}} } {h_{{R_g}{v_{{k_i}}}}}\sqrt {{Q_{{{v'}_{{k_j}}}}}} {h_{{{v'}_{{k_j}}}{R_g}}}{y_{{{v'}_{{k_j}}}}}} \right)}^2}}}{{{{\left( {\sum\limits_{g = 1}^L {\sqrt {{Q_{{R_g}}}} } {h_{{R_g}{v_{{k_i}}}}}{\sigma _{{R_g}}}} \right)}^2} + {{(L\sigma _{{v_{{k_i}}}}^2)}^2}}}.
\end{equation}

When $\sigma_{v_{k_i}}^2 = 0$, (6) simplifies to
\begin{equation}
SNR_{{{v'}_{{k_j}}} \to {v_{{k_i}}}}^{\{ {R_L}\} } = \frac{{{{\left( {\sum\limits_{g = 1}^L {\sqrt {{Q_{{R_g}}}} } {h_{{R_g}{v_{{k_i}}}}}\sqrt {{Q_{{{v'}_{{k_j}}}}}} {h_{{{v'}_{{k_j}}}{R_g}}}{y_{{{v'}_{{k_j}}}}}} \right)}^2}}}{{{{\left( {\sum\limits_{g = 1}^L {\sqrt {{Q_{{R_g}}}} } {h_{{R_g}{v_{{k_i}}}}}{\sigma _{{R_g}}}} \right)}^2}}}.
\end{equation}

On the other hand, when $\sigma_{v_{k_i}}^2 = 0$, the sum of the individual SNRs is given by
\begin{equation}
\sum\limits_{w = 1}^L S NR_{{{v'}_{{k_j}}} \to {v_{{k_i}}}}^{{R_w}} = \sum\limits_{w = 1}^L {\frac{{{{\left| {\sqrt {{Q_{{R_w}}}} {h_{{R_w}{v_{{k_i}}}}}\sqrt {{Q_{{{v'}_{{k_j}}}}}} {h_{{{v'}_{{k_j}}}{R_w}}}{y_{{{v'}_{{k_j}}}}}} \right|}^2}}}{{{{\left| {\sqrt {{Q_{{R_w}}}} {h_{{R_w}{v_{{k_i}}}}}\sigma _{{R_w}}^2} \right|}^2}}}} .
\end{equation}

Let us define
\begin{equation}
\begin{cases}
{U_w} = \left| {\sqrt {{Q_{{R_w}}}} {h_{{R_w}{v_{{k_i}}}}}\sqrt {{Q_{{{v'}_{{k_j}}}}}} {h_{{{v'}_{{k_j}}}{R_w}}}{y_{{{v'}_{{k_j}}}}}} \right|{\rm{ }}\\
{I_w} = \left| {\sqrt {{Q_{{R_w}}}} {h_{{R_w}{v_{{k_i}}}}}\sigma _{{R_w}}^2} \right|.
\end{cases}
\end{equation}

Then (8) can be written as
\begin{equation}
\sum_{w=1}^L SNR_{v'_{k_j} \to v_{k_i}}^{R_w} = \sum_{w=1}^L \frac{U_w^2}{I_w^2}.
\end{equation}

Since $U_w \geq 0$ and $I_w \geq 0$, let $U_w = u_w I_w$, where $u_w$ is a non-negative real number. Thus we have
\begin{equation}
\sum_{w=1}^L SNR_{v'_{k_j} \to v_{k_i}}^{R_w} = \sum_{w=1}^L u_w^2.
\end{equation}

Multiplying the numerator and denominator of (12) by $\left( \sum_{w=1}^L I_w \right)^2$, we have
\begin{equation}
\sum_{w=1}^L SNR_{v'_{k_j} \to v_{k_i}}^{R_w} = \frac{\sum_{w=1}^L u_w^2 \left( \sum_{w=1}^L I_w \right)^2}{\left( \sum_{w=1}^L I_w \right)^2}.
\end{equation}

Expanding $\left( \sum_{w=1}^L I_w \right)^2$, we obtain
\begin{equation}
\left( \sum_{w=1}^L I_w \right)^2 = \sum_{w=1}^L I_w^2 + 2 \sum_{1 \leq i, j \leq L, i \neq j} I_i I_j.
\end{equation}

It is clear that
\begin{equation}
\left( \sum_{w=1}^L I_w \right)^2 \geq \sum_{w=1}^L I_w^2.
\end{equation}

Therefore, we have
\begin{equation}
\sum_{w=1}^L SNR_{v'_{k_j} \to v_{k_i}}^{R_w} \geq \frac{\sum_{w=1}^L u_w^2 \sum_{w=1}^L I_w^2}{\left( \sum_{w=1}^L I_w \right)^2}.
\end{equation}

Applying the Cauchy-Schwarz inequality, we can show that:
\begin{equation}
\sum_{w=1}^L SNR_{v'_{k_j} \to v_{k_i}}^{R_w} \geq \frac{\left( \sum_{w=1}^L u_w I_w \right)^2}{\left( \sum_{w=1}^L I_w \right)^2}.
\end{equation}

Finally, we obtain the following SNR upper:
\begin{equation}
SNR_{v'_{k_j} \to v_{k_i}}^{\{R_L\}} \leq \sum_{w=1}^L SNR_{v'_{k_j} \to v_{k_i}}^{R_w}.
\end{equation}

In real-time cooperative D2D communication, adjusting the transmission strategy of each relay to maximize the D2D signal-to-noise ratio (SNR) yields near-optimal SNR, thereby enhancing communication quality.

\section{Optimal communication in ultra large scale V2X networks}
Furthermore, we evaluate the maximum upper bounds for $SNR_{v'_{k_j} \to v_{k_i}}^{\{R_L\}}$ and $\sum_{w=1}^L SNR_{v'_{k_j} \to v_{k_i}}^{R_w}$, revealing potential optimal channel capacity bounds in 6G cooperative multi-path ultra-massive mobile networks. Additionally, we determine the upper bound of optimal communication when two arbitrary mobile devices are located in specific positions. Moreover, an optimal cooperative strategy can be identified to optimize D2D communication for any two fixed devices in 6G intelligent Internet of Everything networks. According to [11, (2)], the channel capacity is given by:
\begin{equation}
C = \frac{1}{2} \log_2(1 + SNR),
\end{equation}
where $C$ represents the D2D channel capacity, and $SNR$ is the D2D power signal-to-noise ratio. Therefore, the channel capacity of the communication path $v'_{k_j} \to \{R_L\} \to v_{k_i}$ is
\begin{equation}
C_{v'_{k_j} \to v_{k_i}}^{\{R_L\}} = \frac{1}{2} \log_2(1 + SNR_{v'_{k_j} \to v_{k_i}}^{\{R_L\}}).
\end{equation}

\textbf{B-1. Uniform Power Allocation:}
For an arbitrary relay selection subset $\{R_L\}$, if the total power is $Q_{tot}$, and the power of each relay is constant, the channel capacity $C_{v'_{k_j} \to v_{k_i}}^{\{R_L\}}$ is limited by
\begin{equation}
Q_{R_1} = Q_{R_2} = \dots = Q_{R_L} = \frac{Q_{tot}}{L}.
\end{equation}

\textbf{B-2. Optimal Relay Selection with Uniform Power:}
Unlike the arbitrary relay selection in \textbf{B-1}, considering the constraints imposed by relay characteristics (e.g., location, fading coefficients, bit error rate), the D2D channel capacity of each relay may vary. We re-arrange the relays in the subset $\{R_{N_{tot}}\}$ ($\{R_L\} \subseteq \{R_{N_{tot}}\}$) in descending order of their individual channel capacities, denoted as $\{R_{opt-N_{tot}}\} = (R_{opt-1}, R_{opt-2}, \dots, R_{opt-N_{tot}})$, such that:
\begin{equation}
C_{v'_{k_j} \to v_{k_i}}^{R_{opt-1}} \geq C_{v'_{k_j} \to v_{k_i}}^{R_{opt-2}} \geq \dots \geq C_{v'_{k_j} \to v_{k_i}}^{R_{opt-N_{tot}}},
\end{equation}
where
\begin{equation}
C_{v'_{k_j} \to v_{k_i}}^{R_{opt-w}} = \frac{1}{2} \log_2(1 + SNR_{v'_{k_j} \to v_{k_i}}^{R_{opt-w}}).
\end{equation}

We select the $L$ optimal relays from $\{R_{N_{tot}}\}$ to form $\{R_{opt-L}\} = (R_{opt-1}, R_{opt-2}, \dots, R_{opt-L})$, and then compute $C_{v'_{k_j} \to v_{k_i}}^{\{R_{opt-L}\}}$.

\textbf{B-3. Optimal Relay Selection with Optimized Power Allocation:}
Building upon \textbf{B-1} and \textbf{B-2}, cooperative transmission introduces interference due to antennas, duplexing, and mutual interference. Therefore, it is essential to allocate the total power more effectively among the selected optimal relays. For the selected set $\{R^*_{opt-L}\}$, we assume the corresponding power subset is $\{Q^*_{opt-L}\} = (Q^*_{opt-1}, Q^*_{opt-2}, \dots, Q^*_{opt-L})$. Each selected relay requires a minimum power for signal forwarding, denoted as $(Q_1, Q_2, \dots, Q_L)$ ($Q_1 > 0, Q_2 > 0, \dots, Q_L > 0$), such that:
\begin{equation}
C_{v'_{k_j} \to v_{k_i}}^{R^*_{opt-1}} \geq C_{v'_{k_j} \to v_{k_i}}^{R^*_{opt-2}} \geq \dots \geq C_{v'_{k_j} \to v_{k_i}}^{R^*_{opt-L}},
\end{equation}
subject to:
\begin{equation}
Q^*_{opt-1} \geq Q_1, Q^*_{opt-2} \geq Q_2, \dots, Q^*_{opt-L} \geq Q_L,
\end{equation}
and
\begin{equation}
\sum_{w=1}^L Q^*_{opt-w} = Q_{tot}.
\end{equation}

In summary, optimal relay selection and efficient power allocation based on real-time computations of relay features are crucial. For the subset $\{R^*_{opt-L}\}$ with non-uniform power allocation ($Q^*_{R_1} \neq Q^*_{R_2} \neq \dots \neq Q^*_{R_L}$), we have
\begin{equation}
C_{v_{k_i} \to v'_{k_j}}^{\{R^*_{opt-L}\}} \geq C_{v'_{k_j} \to v_{k_i}}^{\{R_{opt-L}\}} \geq C_{v'_{k_j} \to v_{k_i}}^{\{R_L\}}.
\end{equation}

\begin{algorithm}
\caption{Real-time Communication for Multiple Perceptual Vehicles $\left\{ V'_{\text{des}} \right\}$ to Single Vehicle $v_{k_i}$ in 6G Ultra-Large-Scale V2X Networks}
\begin{algorithmic}[1]
\STATE Vehicle $v_{k_i} \in \left\{ V_{\text{sou}} \right\}$ forecasts surrounding traffic areas it will travel through, and identifies a critical vehicle subset $\left\{ V'_{\text{des}} \right\}$ to establish communication.
\FOR{each vehicle $v'_i \in \left\{ V'_{\text{des}} \right\}$}
    \STATE Compute and select optimal $L_i$ cooperative mobile relays with \textbf{B-3} using (23)–(26).
    \STATE Transmit perceptual information to vehicle $v_{k_i}$.
\ENDFOR
\STATE $v'_i$ and $\left\{ V'_{\text{des}} \right\}$ jointly compute to optimize dynamic vehicle-to-vehicle communication over different time intervals.
\end{algorithmic}
\end{algorithm}

\section{simulations}
\begin{figure}[htbp]
  \centering
  \includegraphics[width=3.7 in]{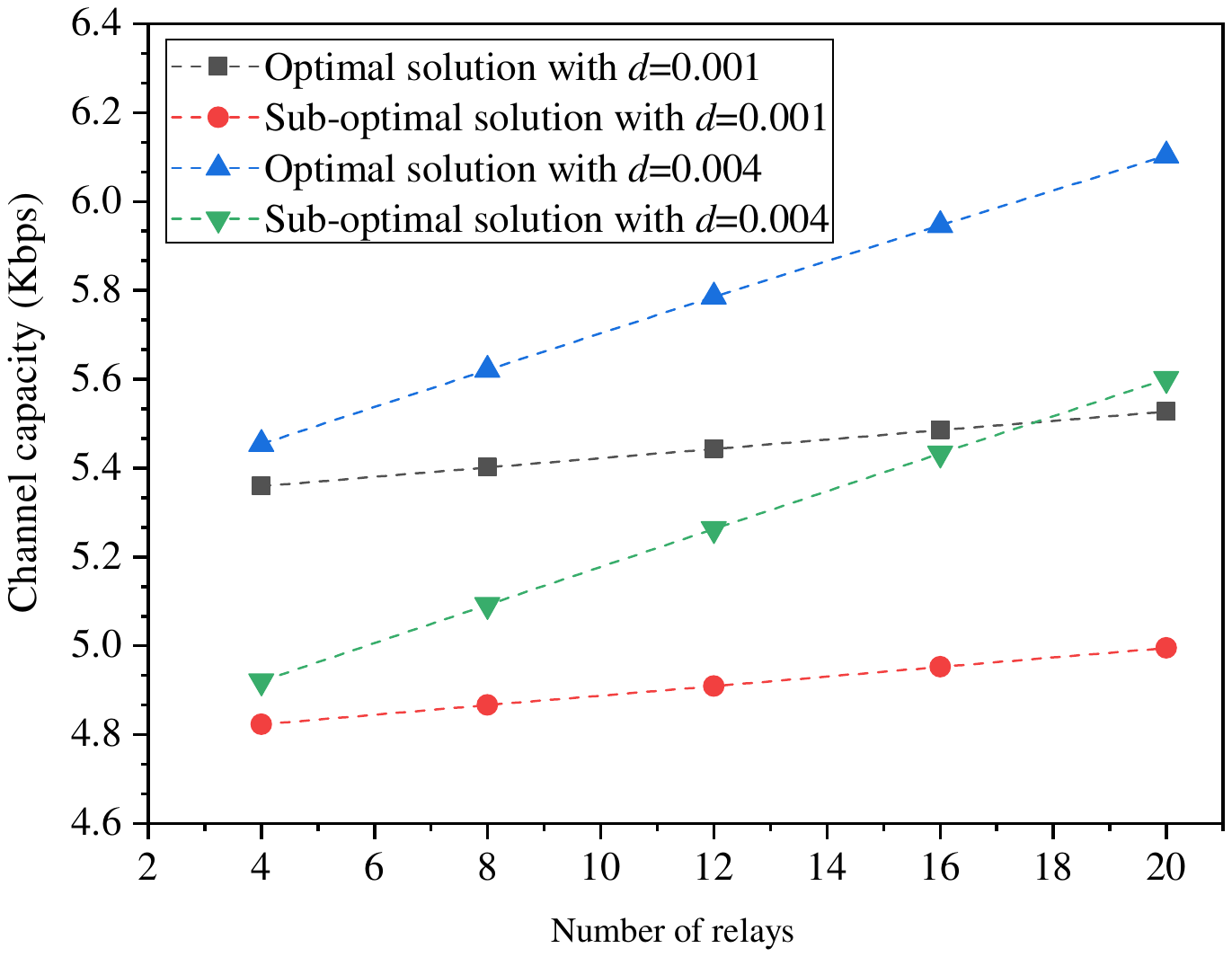}
  \caption{Channel capacity versus number of relays where ${Q_{{\rm{sou}}}} = 18W$}
  \label{Fig.3}
\end{figure}


In this section, we present simulation results for the proposed integrated optimal communication algorithm. Simulation conditions are similar to those in [9]-[14]. To compare our proposed scheme with existing schemes, we assume a bandwidth of 1 kbps for the channel capacity. The simulations aim to demonstrate the advantages of multiple relay selection for vehicle-to-vehicle communication, illustrate the principle of optimal communication in 6G ultra-large-scale V2X networks, and compare our algorithm with existing relay selection algorithms.

\subsection{Optimal Communication with Ultra-Large-Scale Relays}

We first evaluate the optimal communication for vehicle-to-vehicle links based on ultra-large-scale relays. Consider an arbitrary mobile relay $R_g$ (with a total of 100 relays). If $R_g$ moves towards $v'_{k_j}$, we set the channel coefficients $h_{v'_{k_j}R_g} \in [0.8, 0.95]$ and $h_{R_gv_{k_i}} \in [0, 0.65]$. If $R_g$ moves towards $v_{k_i}$, we set $h_{v'_{k_j}R_g} \in [0.75, 0.9]$ and $h_{R_gv_{k_i}} \in [0, 0.7]$. In Figures 3 and 4, we assume $|y_{v'_{k_j}}(t)|^2 = 2$ and $\sigma_{v_{k_i}}^2 = \sigma_{R_g}^2 = 1$. Following the general range of transmit power for each antenna (5W-25W) as reported in [3, 7, 9, 10, 12, 13], we set $Q_{v'_{k_j}} = 15W$ and the average $Q_{R_g} = 20W$ in Figure 3.

In Figure 3, we evaluate the channel capacity versus the number of relays for fading coefficient interval values of $d = 0.001$ and $d = 0.004$. The results show that as the number of relays increases, the channel capacity improves, and a larger $d$ corresponds to a higher channel capacity. This is due to the multi-relay diversity, and appropriate fading coefficient values. Furthermore, we observe that with an average $Q_{R_g} = 20W$, higher transmit rates are achieved with the optimal solution. This is because relays with higher $d$ values are selected.
\begin{figure}[htbp]
    \centering
    \includegraphics[width=3.7in]{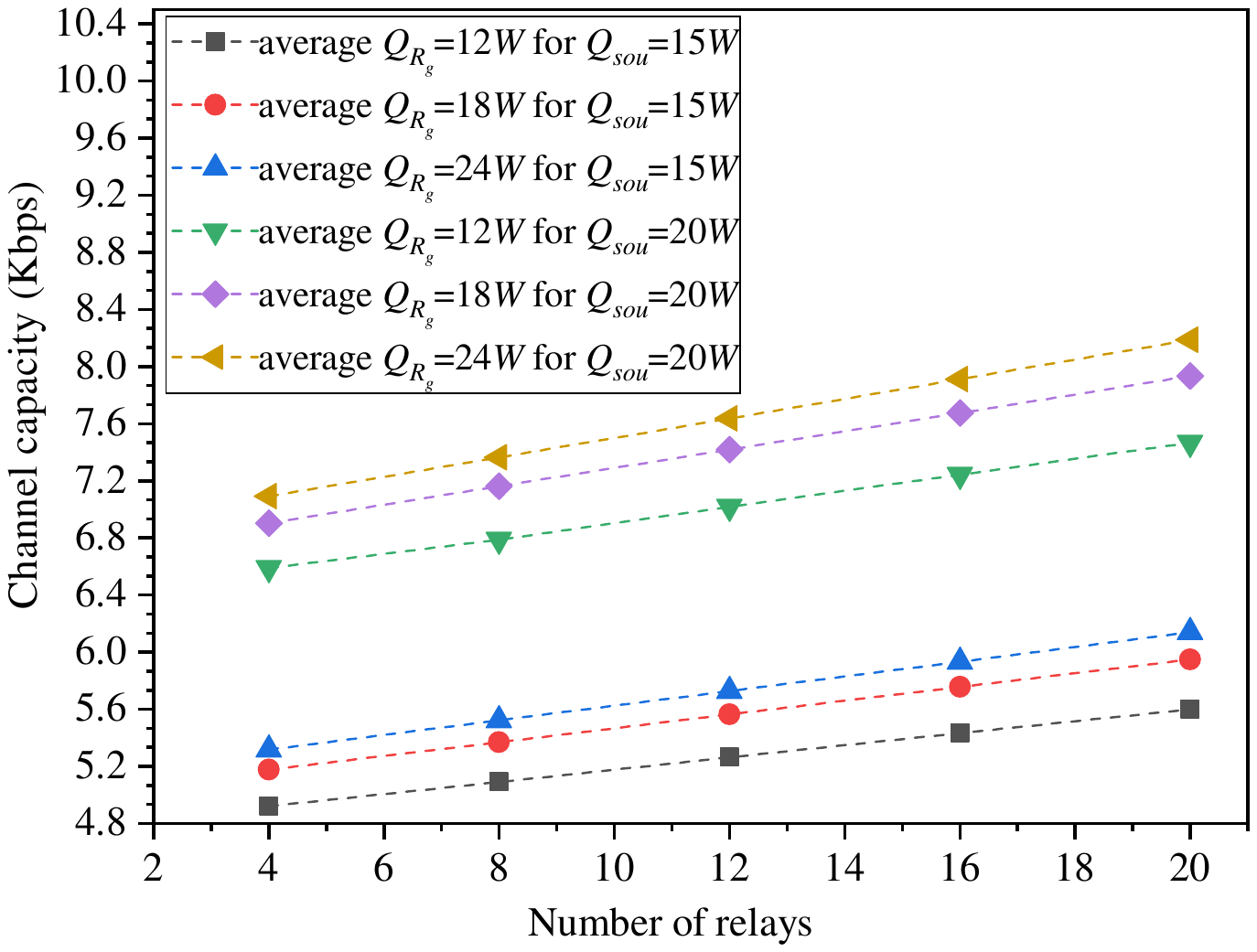}
    \caption{Channel capacity versus number of relays}
    \label{Fig.3}
\end{figure}

Based on the observations in Figure 3, we investigate the channel capacity changes when $Q_{R_g}$ is not constant and $Q_{v'_{k_j}} = 15W, 20W$ in Figure 4. The results indicate that the power value of the source $Q_{v'_{k_j}}$ is more critical than the power value of the relays $Q_{R_g}$. Therefore, one shall prioritize allocating more power to $Q_{v'_{k_j}}$ when selecting optimal relay subsets.

\subsection{Comparison with Existing Algorithms}

Building upon the results in Figures 3 and 4, we further analyze our proposed algorithm and compare it with existing algorithms. Similar to [9, 12, 13, 14], we consider cooperative V2X scenarios with pseudorandom vehicle distribution at a given time slot. We set $Q_{v'_{k_j}} = 18W$, $|y_{v'_{k_j}}|^2 = 2$, $h_{v'_{k_j}R_g} \in [0.5, 0.9]$, $h_{R_gv_{k_i}} \in [0, 0.7]$, $d = 0.004$, $Q_{R_g} \in [10W, 25W]$, $\sigma_{R_g}^2 = 1$ for $g \equiv 1, 2 \pmod{3}$, $\sigma_{R_g}^2 = 0$ for $g \equiv 0 \pmod{3}$, and $\sigma_{v_{k_i}}^2 = 2$. For the algorithm in [12], relays are selected based on higher fading coefficients. For the algorithm in [13], relays are selected based on larger power values with constant and different fading coefficients.

\begin{figure}[htbp]
    \centering
    \includegraphics[width=3.7in]{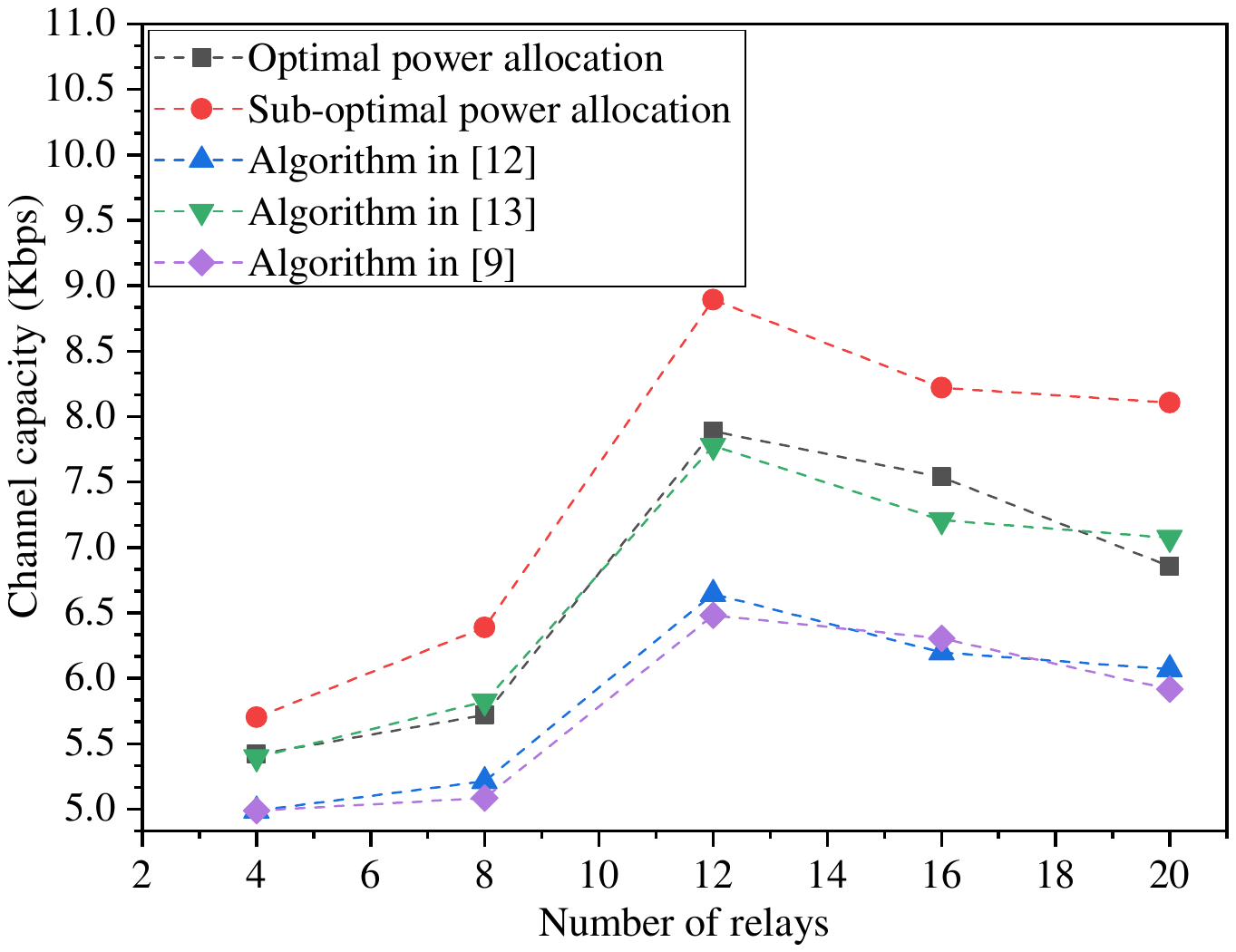}
    \caption{Channel capacity versus number of relays where $Q_{R_g}$ is constant}
    \label{Fig.2}
\end{figure}

Figure 5 shows that our proposed algorithm achieves a 0.8-2.5 kbps higher channel capacity compared to that in [9]-[14]. This is because we comprehensively consider various relay factors, compute optimal and sub-optimal channel capacities, and select better relays from varied vehicle subsets. In contrast, the algorithms in [9, 12, 13] only consider one or partial relay factors and do not present an optimal power allocation scheme. Interestingly, we observe that increasing relay diversity does not always yield a higher channel capacity. In Figure 5, the channel capacity decreases when the number of relays exceeds 12. This is because the noise from some relays has a greater impact when signals are combined at the receiver $v_{k_i}$, resulting in a smaller total SNR. The descending rate also varies among different algorithms.

\section{Conclusions}
In this study, we have studied optimal real-time communication for 6G ultra large scale V2X networks. Compared with existed cooperative vehicular algorithms, the proposed algorithm leads to higher D2D channel capacity. Besides, we have analyzed the principle of 6G D2D optimal real-time communication based on ultra large scale relays selection, and proved the existence of upper bound of channel capacity.

\section*{Acknowledgment}
This work was supported in part by Fundamental Research Funds for the Central Universities of China (Grant No.PA2024GDSK0114 and No.JZ2024HGTG0311),  National Natural Science Foundation of China (Grant No.62371180), Open Research Project of the State Key Laboratory of Industrial Control Technology, China (Grant No. ICT2025B23), Sichuan Science and Technology Program (Grant No.2025HJRC0023), Research Fund of Talent Introduction of Chengdu University of Information and Technology (Grant No.376613).

\bibliographystyle{ieeetr}

\end{document}